\documentclass[aps,prl,twocolumn,superscriptaddress,floatfix,showpacs]{revtex4}
\usepackage{graphicx}
\usepackage{amssymb}
\usepackage{amsmath}

\newcommand{\ssc}{\scriptscriptstyle}
\renewcommand{\phi}{\varphi}

\begin{document}

\title{The Amplitude Mode in the Quantum Phase Model}
\author{S.D. Huber}
\affiliation{Theoretische Physik, ETH Zurich, CH-8093
Zurich, Switzerland}
\author{B. Theiler}
\affiliation{Theoretische Physik, ETH Zurich, CH-8093
Zurich, Switzerland}
\author{E. Altman}
\affiliation{Department of Condensed Matter Physics, The Weizmann
Institute of Science, Rehovot, 76100, Israel}
\author{G. Blatter}
\affiliation{Theoretische Physik, ETH Zurich, CH-8093
Zurich, Switzerland}

\date{\today}

\begin{abstract}
We derive the collective low energy excitations of the quantum phase model
of interacting lattice bosons within the superfluid state using a dynamical
variational approach. We recover the well known sound (or Goldstone) mode
and derive a gapped (Higgs type) mode that was overlooked in previous
studies of the quantum phase model.  This mode is relevant to ultracold
atoms in a strong optical lattice potential. We predict the signature of
the gapped mode in lattice modulation experiments and show how it evolves
with increasing interaction strength.
\end{abstract}


\maketitle

Interactions can have a dramatic influence on the properties of superfluids
at low temperatures. In the most extreme case, such as lattice bosons at
commensurate filling, interactions drive a quantum phase transition to an
insulating phase \cite{fisher89}. But even inside the superfluid phase,
interactions may greatly impact on basic properties such as the excitations,
with the roton minimum in the spectrum of Helium providing a well known
example. In this letter, we investigate the effect of increasing interaction
strength on the collective modes in a superfluid of lattice bosons; the
latter are usually modeled by the Bose-Hubbard model incorporating both
hopping (parameter $t$) and local interactions ($U$).  The question has
been brought into focus by experiments with ultracold atoms in optical
lattices that were able to control the interaction strength and even
drive a transition to the insulating state \cite{Greiner02}. More recent
experiments have probed the excitation spectrum of the superfluid in
this strongly correlated regime by measuring the energy absorption rate
in response to periodic lattice modulations. The interpretation of the
lattice modulation experiments is complicated by the large magnitude of the
perturbation and the presence of a confining potential \cite{Kraemer04};
nevertheless, the results are very suggestive of the existence of a gapped
collective mode in the superfluid phase \cite{Cazalilla06,Huber07}, which
is at the center of interest in the present work.
\begin{figure}[!bt] \includegraphics{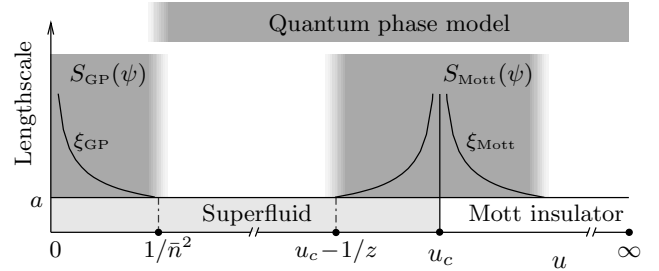} \caption{
Schematic overview of models describing the superfluid phase of lattice
Bosons. For small (dimensionless) interactions $u=U/2t\bar n z<1/\bar n^{2}$,
the Bose-Hubbard model is well approximated by the continuum Gross-Pitaevskii
action $S_{\ssc {\rm GP}}(\psi)$. Close to the critical interaction
$u \approx u_{c} = 2$, the dynamics is described by the critical theory
$S_{\ssc {\rm Mott}}(\psi)$. For $u> 1/\bar n^{2}$, the Bose-Hubbard model
is equivalent to the quantum phase model, analyzed in detail in this work.
}
\label{fig:overview}
\end{figure}

The Bose-Hubbard model generates two {\it effective} low-energy field theories
for the complex order-parameter field $\psi$, see Fig.\ \ref{fig:overview}:
for weak interaction $U\bar n \ll t$, the effect of the lattice can be
absorbed in an effective band mass and one arrives at the Galilean-invariant
Gross-Pitaevskii theory (nonlinear Schr\"odinger equation; $\bar n$ is the
mean filling per site). Quite remarkably, close to the superfluid-insulator
transition at $U \sim t \bar n$, the combined action of the lattice and the
interaction leads to a Lorentz-invariant critical theory (nonlinear Klein
Gordon equation) \cite{Sachdev99}. The first-order time derivative in the
Gross-Pitaevskii theory defines the density $\rho = |\psi|^2$ and hence any
density mode is bound to the phase degree of freedom, resulting in the unique
and well known sound (or Goldstone) mode.  This differs from the
Lorentz-invariant critical theory, where the second-order time derivative
spoils the relation between the order-parameter modulus and the density; as a
consequence, this theory admits the possibility of independent amplitude
(Higgs) and phase (Goldstone) modes. Naturally, the question poses itself, how
the Higgs mode emerges upon increasing the interaction $U$. This question
should be addressed within a {\it microscopic} theory that explicitly accounts
for the interplay between interactions and the lattice, as is done by the
quantum phase model \cite{Doniach81,Simanek80}, a suitable approximation of
the Bose-Hubbard model for large filling $\bar n$.  In what follows, we
discuss in more detail the two effective low-energy theories for the
Bose-Hubbard model and their interrelation with the quantum phase model and
then concentrate on the derivation of the amplitude mode and its experimental
observation.

We start with the Bose-Hubbard model in a coherent state formulation where
the action is given by
\begin{align}
\nonumber 
   S_{\ssc {\rm BH}}(\{\psi_i\}) 
   &= \!\int_{0}^{\hbar\beta}\!\!\!\!\!\! d\tau \,
     \mathcal{L}_{\ssc {\rm BH}}(\{\psi_i\})
      = \!\int_{0}^{\hbar\beta}\!\!\!\!\!\! d\tau \,\biggl[\sum_{i}
      \psi_{i}^{*}(\hbar\partial_{\tau} \!-\! \mu)\psi_{i} \\
   &~~- t \sum_{\langle i,j \rangle} \psi_{i}^{*} \psi_{j} 
   + \frac{U}{2} \sum_{i} |\psi_{i}|^{2}(|\psi_{i}|^{2}-1)\biggr];
   \label{eqn:bhm}
\end{align}
here, $\mu$ is the chemical potential and  $\beta=1/k_{\rm B}T$ is the
inverse temperature.  The bosonic fields $\psi_{i}$ denote the amplitude
of particles in a Wannier state at site $i$.

The diverging correlation length $\xi_{\ssc {\rm Mott}}\propto 1
/\sqrt{z(u_{c}-u)}$ near the (commensurate) superfluid--Mott-insulator
transition at $u=U/2t\bar n z=2$ allows for a description in terms of a
continuum critical theory ${\mathcal L}_{\ssc {\rm Mott}} (\psi) =
|\hbar\partial_{\tau}\psi|^2/Jz + Jz(u_c-u)[\xi_{\ssc {\rm Mott}}^2
|\nabla\psi|^2 - |\psi|^2] +U|\psi|^4/2$, with $z=2d$ the coordination number
and $J=2t\bar n$.  The emergent (Higgs) mode involves a collective
oscillation of the amplitude $|\psi|$ of the order parameter with a frequency
that vanishes at the transition. Such oscillation of the order parameter is
accompanied by a local change in the non-condensed fraction, leaving the local
density unchanged. Furthermore, this mode is independent of the usual sound
mode; this can be understood from the fact that the order parameter vanishes
towards the transition while the density remains constant.

Clearly the gapped amplitude mode is absent in the opposite regime, of very
weak interactions $u\ll 1/\bar n^{2}$. In this case, the length scale set by
the interaction is the well known healing length $\xi_{\ssc {\rm
GP}}=a\sqrt{t/U \bar n} \gg a$. The fact that $\xi_{\ssc {\rm GP}}$ is much
larger than the lattice constant $a$ allows for a continuum (Gross-Pitaevskii)
description ${\mathcal L}_{\ssc {\rm GP}}(\psi) = \psi^{*} \hbar
\partial_{\tau}\psi+ U\bar n \,\xi_{\ssc {\rm GP}}^2 |\nabla \psi|^{2} +
U|\psi|^{4}/2$. The effective Galilean invariance ensures that at zero
temperature there is no amplitude mode independent of first sound.

Comparing the regions of validity of the above coarse grained theories, see
Fig.~\ref{fig:overview}, we observe that these regimes are parametrically
separated from each other for $\bar n \gg 1$.  The absence of a diverging
length scale in the intermediate regime $1/\bar n^{2}<u< 1$ renders the effect
of the lattice relevant. As the Galilean invariance is explicitly broken, the
existence of an amplitude mode cannot be ruled out.  For large site occupancy,
the Hubbard model in the intermediate (and strong) interaction regimes is
equivalent to the (simpler) quantum phase model \cite{Doniach81,Simanek80}
\begin{equation}
\label{eqn:qpm}
\hat H_{\ssc {\rm QPM}} = -J \sum_{\langle i,j \rangle}
\cos(\hat\phi_{i}-\hat\phi_{j}) + \frac{U}{2} \sum_{i} \delta\hat n_{i}^{2},
\end{equation}
with the Josephson coupling $J=2t\bar n$. The conjugate operators
$\hat\phi_{i}$ and $\delta\hat n_{j}$, $[\hat\phi_{i},\delta\hat
n_{j}]=i\hbar \delta_{ij}$, describe the local phase and deviation from
mean filling, respectively. The derivation of $\hat H_{\ssc {\rm QPM}}$ from
(\ref{eqn:bhm}) involves an integration over density fluctuations under the
assumption $\langle \delta \hat n \rangle / \bar n \ll 1$ \cite{fisher89},
which is valid for $u\!\approx\! 1/\bar n^2$, and subsequent (re-)
quantization. Here we restrict our considerations to the case of integer
filling $\bar n \in \mathbb N$.

We analyze the quantum phase model within a dynamical variational approach
\cite{Jackiw79}, which accounts for both phase and amplitude degrees of
freedom and allows us to capture the low-energy physics of a depleted
condensate near the Mott-insulator transition. We first derive the static
properties in a mean-field approach and  then include dynamics within a
Gaussian approximation. Finally, we discuss the response of the system to
an external lattice modulation, thereby connecting our findings with recent
experiments \cite{Schori04}.  Before proceeding, we note that a gapped
excitation closely related to the one considered here has been identified
by Cazalilla {\it et al.} \cite{Cazalilla06} using a bosonization approach
to a system with a strongly anisotropic optical lattice potential.

Our variational wave function has the Gutzwiller form
\begin{align}
\nonumber
|\Psi\rangle &= \prod_{i} \sum_{n_{i}} 
f_{n_{i}}(\sigma_{i},\phi_{i})
| n_{i} \rangle
\quad\mbox{with} \\
\label{eqn:static-wavefunction}
f_{n_{i}}(\sigma_{i},\phi_{i}) &= 
\frac{1}{(2\pi \sigma_{i})^{1/4}}
e^{-(n_i-\bar n)^2/2\sigma_i} e^{i\bar n \phi_i},
\end{align}
where $|n_{i}\rangle$ is a particle-number state of the
quantum phase model and $\delta\hat n_{i} |n_{i}\rangle=
(n_{i}-\bar n)|n_{i}\rangle $. The order-parameter in the state
given by (\ref{eqn:static-wavefunction}) is $\psi_{i}=\langle
\exp(-i\hat\phi_i)\rangle=e^{-1/4\sigma_{i}}e^{-i\phi_{i}}$, so that
$\phi_{i}$ and $\sigma_{i}$ determine phase and amplitude (fluctuations)
of $\psi_{i}$, respectively. The wave-function $|\Psi\rangle$ has a norm
$\langle \Psi | \Psi \rangle =\prod_{i}\sum_{n_{i}}f_{n_{i}}^{*}f_{n_{i}}
=\prod_{i} \sum_{m}\exp[-(m\pi \sigma_{i})^{2}]$ and is not properly
normalized, a consequence of the discreteness of the particle numbers.
This is not a problem as long as we stay away from the transition so
that the particle number fluctuation is large ($\sigma_i\gg1$). We will
concentrate on this regime.

\begin{figure}[!t]
\includegraphics{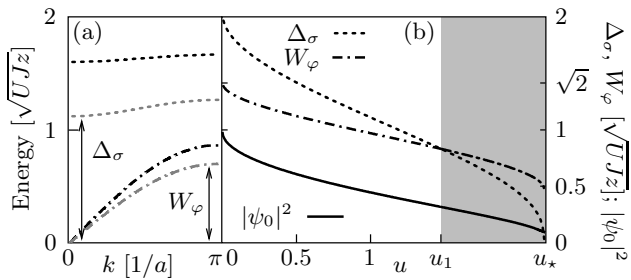}
\caption{(a) Spectra for the phonon (dash-dotted line) and the amplitude
mode (dotted line). Black curves correspond to $u=U/Jz=0.25$, gray curves to
$u=1$. (b) Gap $\Delta_{\sigma}$ of the amplitude mode (dotted line) and the
phonon bandwidth $W_{\phi}$ (dash-dotted line) vs dimensionless coupling
$u= U/Jz$ in the superfluid phase. The gray shaded area corresponds to the
Mott phase. The solid line denotes the condensate fraction $|\psi_{0}|^{2}$.}
\label{fig:disp}
\end{figure}

The variational energy $\epsilon_{\rm var}=\langle \Psi| \hat H_{\rm QPM}
|\Psi \rangle$ is given by
\begin{equation}
\epsilon_{\rm var} =
-J \sum_{\langle i,j \rangle} 
e^{-(\sigma_{i}^{-1}+\sigma_{j}^{-1})/4}
\cos(\phi_{i}-\phi_{j}) +\frac{U}{4}\sum_{i} \sigma_{i}
\end{equation}
and is minimized for $\phi_i = \phi_{\ssc{\rm mf}}\equiv 0$ and $\sigma_i
= \sigma_{\ssc{\rm mf}} \equiv 0$ in the Mott phase. In the superfluid
phase $\sigma_{\ssc{\rm mf}}\neq 0$, leading to an order-parameter (or
condensate fraction)
\begin{equation}
\label{eqn:static-psi0}
|\psi_{0}|^{2} \equiv e^{-1/2\sigma_{\ssc{\rm mf}}}
=e^{2W\bigl(-\sqrt{u/16}\bigr)},
\end{equation}
where $W(x)$ is the Lambert-$W$ function \cite{comment:lambertw}, cf.
Fig.~\ref{fig:disp}(b).

If this scheme is carried out blindly all the way to strong coupling, the
transition to the Mott insulator appears as of first (rather than second)
order.  Besides a local minimum at $\sigma_{i}=0$, which is always present,
a second minimum first appears at $u_{\star} = 16/e^{2} \approx 2.16$
(spinodal point). This spurious first order transition (at $u_{1}\approx
1.47$) is due to the failure of the variational wave-function when the
particle number fluctuations become small $\langle \delta \hat n_i^2\rangle
\ll  1$ near the transition \cite{Simanek80}. However, as mentioned earlier
we will use this approach only away from the critical regime of ${\mathcal
L}_{\ssc {\rm Mott}}(\psi)$ where (\ref{eqn:static-wavefunction}) should
be a good approximation.

Next, we describe the fluctuations above the mean-field ground state
(\ref{eqn:static-wavefunction}) using a time-dependent variational principle.
Following Ref.\ \cite{Jackiw79} we aim at an effective action (we switch
to real time)
\begin{equation}
\label{eqn:effective-action}
S_{\ssc {\rm eff}} = 
\int dt\, {\mathcal L}_{\ssc {\rm eff}} = 
\int dt\, 
\langle \Psi | i\hbar\partial_{t} - \hat H_{\rm QPM} | \Psi \rangle.
\end{equation}
While the minimizer of this expression provides the exact action in case
of unrestricted and properly normalized variational states $|\Psi\rangle$,
here, we restrict ourselves to the class of variational wavefunctions
(cf.\ \ref{eqn:static-wavefunction})
\begin{equation}
\label{eqn:dynamic-wavefunction}
f_{n_{i}} = \frac{1}{(2\pi \sigma_{i})^{1/4}}
e^{-(n_{i}-\Phi_{i})^{2}\bigl(1/2 \sigma_{i}-2i\Sigma_{i}\bigr)}
e^{i (n_{i}-\Phi_{i})\phi_{i}}.
\end{equation}
This ansatz generates the Lagrangian
\begin{align}
\lefteqn{
   {\mathcal L}_{\ssc {\rm eff}} =
   \sum_{i} \hbar \dot\sigma_{i}\Sigma_{i} + \hbar \dot\phi_{i}\Phi_{i}
   -\frac{U}{4}\sigma_{i} - \frac{U}{2}(\bar n - \Phi_{i})^{2}}
    \label{eqn:full-lagrangian}
     \\
&       \nonumber
   +J \sum_{\langle i,j \rangle}
   e^{-(1+16\Sigma_i^2\sigma_i^2)/4\sigma_i}
   e^{-(1+16 \Sigma_j^2\sigma_j^2)/4\sigma_j}
   \cos(\phi_{i}-\phi_{j}).
\end{align}
The new parameters $\Sigma_{i}$ and $\Phi_{i}$ in the wave function
(\ref{eqn:dynamic-wavefunction}) assume the role of canonically conjugate
fields and allow the order parameter's amplitude ($\sigma_{i}$) and phase
($\phi_{i}$) degrees of freedom to acquire dynamics.  The interaction terms
$\propto U$ are linear/quadratic in these fields and the coupling between
phase ($\varphi_i$) and amplitude ($\sigma_i$) degrees of freedom is only
through the hopping term $\propto J$.

We obtain an effective action in terms of the fields $\sigma_{i}$ and
$\phi_{i}$ by solving the Euler-Lagrange equations for $\Sigma_{i}$
and $\Phi_{i}$.  The static solution of the Euler-Lagrange equations for
the $\sigma_{i}$ and $\phi_{i}$ fields reproduces the mean-field ground
state Eq.~(\ref{eqn:static-psi0}). Expanding the effective action around
($\sigma_{\ssc{\rm mf}}$, $\phi_{\ssc{\rm mf}}$) to second order, we find
decoupled amplitude and phase degrees of freedom in ${\mathcal L}_{\ssc
{\rm eff}}$. We use a Legendre transformation [$(x_{i},\dot x_{i})
\rightarrow (x_{i},\Pi_{x_{i}})$, where  $x=\sigma$, $\phi$] to
quantize the real, classical fields introducing ladder operators ($[\hat
x_{\bf k},\hat x_{{\bf k}'}^{\dag}]=\delta_{{\bf k},{\bf k}'}$):
\begin{equation}
\left(
\begin{matrix}
\hat x_{i} \\ \hat\Pi_{x_{i}}
\end{matrix}
\right)
= \frac{1}{\sqrt{N}} \sum_{\bf k}
\left(
\begin{matrix}
A^{x}_{\bf k} \\ iB^{x}_{\bf k}
\end{matrix}
\right)
\biggl[
\hat x_{\bf k}e^{i{\bf k}\cdot {\bf r}_{i}} \pm
\hat x_{\bf k}^{\dag}e^{-i{\bf k}\cdot {\bf r}_{i}}
\biggr].  
\label{eqn:quantization}
\end{equation}
Here,  $A_{\bf k}^{x} = \sqrt{\hbar/2 m_{x} \omega_{x}({\bf k})}=B_{\bf
k}^{x}/\omega_{x}({\bf k})$, $N$ is the number of sites, and the``masses''
are given by $m_{\sigma}=\hbar^{2}|\log|\psi_{0}||/2Jz|\psi_{0}|^{2}$
and $m_{\phi}=\hbar^{2}/U$.  We finally obtain the quadratic Hamiltonian
\begin{equation}
\label{eqn:effective-hamiltonian}
\hat H_{\rm eff}= 
\sum_{\bf k} 
\hbar \omega_{\phi}({\bf k}) \hat\phi_{\bf k}^{\dag}\hat \phi_{\bf k}
+\sum_{\bf k}
\hbar \omega_{\sigma}({\bf k}) \hat\sigma_{\bf k}^{\dag} 
\hat\sigma_{\bf k}
\end{equation}
describing phase and amplitude fluctuations above the mean-field ground
state. The dispersions of the modes are given by
\begin{align}
\nonumber
\hbar \omega_{\sigma}({\bf k}) & = 
\sqrt{UJz}\, |\psi_{0}|^{2}
\\ 
\nonumber
& \times \sqrt{32 |\log|\psi_{0}||^{2}
\{2+\log|\psi_{0}|   [1+\gamma({\bf k})]\}/u},
\\
\label{eqn:spectra}
\hbar \omega_{\phi}({\bf k}) & =
\sqrt{UJz}\, |\psi_{0}| \sqrt{1-\gamma({\bf k})},
\end{align}
where $\gamma({\bf k})=(2/z)\sum_{l=1}^{d}\cos({\bf k}\cdot{\bf a}_{l})$
with lattice vectors ${\bf a}_{l}$. The amplitude mode is characterized
by a finite gap $\Delta_{\sigma} =\hbar\omega_{\sigma}(0)$ extending
throughout the entire range of parameters \cite{comment:gap_vanish},
cf.\ Fig. \ref{fig:disp}; for small $u\rightarrow0$ the amplitude mode
becomes nondispersive, i.e, $\hbar\omega_{\sigma}({\bf k}) \approx
\Delta_{\sigma}$ for all ${\bf k}$. The phase mode, on the other hand,
is gap-less and characterized by a sound velocity $v_{s}^{\ssc {\rm
eff}}=\sqrt{UJz|\psi_{0}|^{2}}  a/\hbar$. This is the Gross-Pitaevskii
result, up to the factor $|\psi_{0}|^{2}$ accounting for the depletion of
the condensate at large $u$.

Because the sound mode corresponds to a density fluctuation it can be probed
directly by measuring the  dynamic structure factor related to the
density-density response of the system. This has been done in systems of
ultracold atoms by using Bragg spectroscopy \cite{Stenger99}.  The gapped
amplitude mode described above is not directly accessible to Bragg
spectroscopy because it does not involve a density modulation. Rather, it is
excited by perturbations that act to modulate the particle number variance, or
equivalently, the distance to the Mott insulator phase.  Experiments in
ultracold atoms have done just that \cite{Stoferle04}. By modulating the
strength of the optical lattice potential, those experiments effectively
modulate the tunneling which is exponentially sensitive to the lattice depth.
We remark that a similar modulation of the coupling was proposed to
be relevant in experiments on Josephson junction arrays measuring the
attenuation of ultrasound \cite{Muhlschlegel84}; however, only coupling to the
phase degrees of freedom was considered in this context.

\begin{figure}[t]
\includegraphics{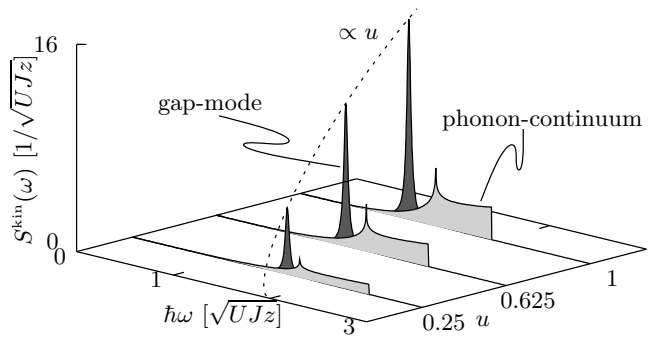}
\caption{
The response function to the lattice modulation $S^{\ssc {\rm kin}}(\omega)$
within the quadratic theory for different values of the interaction
strength $u$ in two dimensions. The dark gray peak relates to the
absorption of a single amplitude mode at $\hbar\omega=\Delta_{\sigma}$
(for illustrational purposes we give the delta-peak a finite width). The
area in faint gray corresponds to the two-phonon continuum excited at
energies below twice the phonon bandwidth $\hbar\omega \leq 2 W_{\phi}$
(the logarithmic divergence is a density-of-states effect peculiar to two
dimensions). Both absorption probabilities scale as $u$ at small $u$.
}
\label{fig:skin}
\end{figure}

To calculate the response to a lattice modulation, we return to
the classical theory ${\mathcal L}_{\ssc {\rm eff}}$, cf.\ Eq.\
(\ref{eqn:full-lagrangian}), and extract the kinetic energy by replacing
$J\rightarrow J + h$, keeping track of terms $\propto h$. Going through
the re-quantization procedure (\ref{eqn:quantization}), we obtain the
kinetic-energy operator $\hat T = \hat T_{\sigma} + \hat  T_{\phi}$ which
we expand to lowest order in $u$
\begin{align}
\label{eqn:mainresult}
\hat T_{\sigma} &\approx 
\sqrt{u}\;\;  \frac{z}{2\sqrt{2}}\sqrt{N} (\hat \sigma_{0}+ \hat
\sigma_{0}^{\dag}),
\\  
\nonumber
\hat T_{\phi} &\approx
\sqrt{u} \;\;  \sum_{\bf k} \frac{z}{4} \sqrt{1-\gamma({\bf k})}
[
\hat \phi_{\bf k} \hat\phi_{-{\bf k}}+
\hat \phi^{\dag}_{\bf k} \hat\phi^{\dag}_{-{\bf k}}+
\hat \phi^{\dag}_{\bf k} \hat\phi_{\bf k}
].
\end{align}
The operator $\hat T_{\sigma}$ describes the expected direct coupling of the
lattice modulations to the amplitude mode at ${\bf k}=0$. The other term $\hat
T_{\phi}$ describes pair-excitations of phase modes (phonons) by the lattice
modulations. Both operators scale as $\sqrt{u}$ for small interactions.

Accordingly, the linear response function $S^{\ssc{\rm
kin}}(\omega)=\sum_{n}|\langle n| \hat T|0\rangle|^2 \delta(\hbar
\omega-\hbar\omega_{n0})$, with $|n \rangle$ the eigenstates of
the unperturbed Hamiltonian (\ref{eqn:effective-hamiltonian}) and
$\hbar\omega_{n0}$ the energy differences between states $|0\rangle$ and
$|n\rangle$, is composed of two parts including a single mode peak due to
the gapped mode and a two-phonon continuum:
\begin{equation}
\nonumber
S^{\ssc {\rm kin}}(\omega) 
= \frac{2 z^{2}|\psi_{0}|^{4}|\log|\psi_{0}||^{2}}{\sqrt{1+\log|\psi_{0}|}} 
N \delta(\hbar \omega - \Delta_{\sigma})
+ S^{\ssc {\rm kin}}_{\ssc {\rm 2p}}(\omega).
\label{eqn:skin}
\end{equation}
Both terms  scale as $u$, see Fig.~\ref{fig:skin}. In two
dimensions, the two-phonon continuum is given by $S^{\ssc {\ssc {\rm
kin}}}_{\ssc {\rm 2p}}(\omega)\propto u \omega^{3} K (\hbar\omega
\sqrt{1-(\hbar\omega/2W_{\phi})^{2}}/W_{\phi})$, with $K(x)$ the complete
elliptic function; going to higher dimensions requires numerical evaluation.

Expanding the effective Lagrangian ${\mathcal L}_{\ssc {\rm eff}}$ to third
order in the fields $\sigma_{i}$ and $\phi_{i}$ provides us with the most
relevant decay channel of the amplitude mode which turns out to involve two
counter-propagating phonons. We find that the small (in dimensions larger than
1) phase space for such a process leaves the mode under-damped and damping
even becomes irrelevant in the limit $u\rightarrow0$.  Finally, we comment
that the same results can be obtained using an RPA type calculation; such an
approach allows for a systematic improvement of the above results and will be
discussed in a future publication \cite{Huber07b}.

In summary, we derived a gapped amplitude (Higgs-type) mode showing up in the
quantum phase model using a dynamical variational approach and discussed its
relevance in the context of superfluid bosonic atoms exhibiting strong
correlations due to the presence of an optical lattice.  Our analysis
demonstrates that this mode persists down to weak coupling $u \approx 1/\bar
n^2$ where the Gross-Pitaevskii description takes over.  We note that broken
translation invariance (due to the presence of a lattice) is crucial for the
existence of this mode. This is also reflected in the experiment where the
coupling to this mode is introduced through a modulation of the lattice; while
experiments have provided evidence for the presence of such a mode
\cite{Schori04}, their spectral resolution does not yet allow for its detailed
analysis. Coupling to this mode in a Josephson junction array seems difficult
due to the rigidity of the coupling parameters; on the other hand, the
presence of a charge density wave in NbSe$_2$ may give access to this mode
\cite{Sooryakumar80,Lei85,Varma02}. While our analysis applies to the case of
commensurate filling, we expect our results to remain valid away from this
limit as the presence of the amplitude mode is connected with a squeezed
Gaussian wave function and does not require the presence of particle-hole
symmetry.

We thank F.\ Hassler and R.\ Barankov for insightful discussions and
acknowledge financial support from the Swiss National Foundation through
the NCCR MaNEP. S.H.\ thanks the Institut Henri Poincare - Centre Emile
Borel for hospitality and support. E.A. was supported by the US-Israel 
binational science foundation.

\bibliographystyle{apsrev}
\bibliography{ref,comments}

\begin{thebibliography}{19}
\bibitem[{com({\natexlab{a}})}]{comment:lambertw}
\bibinfo{note}{{The Lambert-$W$ function is defined as the solution to
  $W(x)\exp[W(x)]=x$. It is real, negative, and monotonous for $x \in[-1/e,0]$
  with $W(-1/e) = -1$ and $W(0)=0$. Furthermore, W$(x)\approx x$ for
  $x\rightarrow0$.}}
\end{thebibliography}

\end{document}